\documentclass[prd,reprint,10pt]{revtex4-1}

\usepackage{fullpage}
\usepackage{amsmath, amsthm, amsfonts, amssymb, mathtools, calrsfs, tensor, tikz-cd}
\usepackage[mathscr]{euscript}
\usepackage{graphicx}
\graphicspath{ {images/} }
\usepackage{enumitem}
\setlist[description]{font=\normalfont}

\newcommand*\diff{\mathop{}\!\mathrm{d}}

\newcommand*\scr[1]{\mathscr{#1}}

\newcommand*\R{\mathbb R}

\newcommand*\te[1]{\text{#1}}

\newcommand*\p[1]{\left(#1\right)}
\newcommand*\ps[1]{\left[#1\right]}
\newcommand*\pc[1]{\left\{#1\right\}}
\newcommand*\f[2]{\frac{#1}{#2}}

\newcommand*\I{\te{i}}

\newcommand*\td[3]{\frac{d^{#3}#1}{d #2^{#3}}}

\usetikzlibrary{matrix,arrows,decorations.pathmorphing}

\begin{document}

\title{Reconsidering seismological constraints\\ on the available parameter space of macroscopic dark matter}
\author{David Cyncynates, Joshua Chiel, Jagjit Sidhu and Glenn D. Starkman}
\affiliation{Physics Department/CERCA/ISO\\ Case Western Reserve University\\ Cleveland, Ohio 44106-7079, USA}

\begin{abstract}
Using lunar seismological data, constraints have been proposed on 
the available parameter space of macroscopic dark matter (macros).
We show that actual limits are considerably weaker 
by considering in greater detail 
the mechanism through which macro impacts generate detectable seismic waves,
which have wavelengths considerably longer than the diameter of the macro.
We show that the portion of the macro parameter space that can be ruled out 
by current seismological evidence is considerably smaller than previously reported,
and specifically that candidates with greater than or equal to nuclear density
are not excluded by lunar seismology.
\end{abstract}
\maketitle
\section{Introduction}

If General Relativity is correct, 
then dark matter constitutes most of the mass density of the Galaxy. 
Yet,  decades after the case for dark matter became compelling \cite{Rubin:1980zd}
and widely accepted (although see \cite{Lelli:2016zqa})
we still do not know what it is. 
The most widely considered and searched for candidates 
are new particles not found in the Standard Model of particle physics,
such as the generic class of  Weakly Interacting Massive Particles (WIMPs) 
(especially the Lightest Supersymmetric Particle)
and axions. 

In this paper, 
we consider instead a class of macroscopic dark matter (macros) candidates.
The theoretical motivation for this stems originally from 
the work of Witten \cite{Witten:1984rs},
and later, more carefully Lynn, Nelson and Tetradis \cite{Lynn:1989xb}.
Macroscopic objects made of baryonic matter with sizable \lq\lq{strangeness}\rq\rq
(i.e. many of the valence quarks are strange quarks, rather than the usual up and down
quarks found in protons and neutrons)
may be stable, 
and may have been formed before nucleosynthesis \cite{Witten:1984rs,Lynn:1989xb}, 
thus evading the principal constraint on baryonic dark matter.
The  appeal of such a dark matter candidate 
is that there would be no need to invoke the existence of new particles 
to explain the observed discrepancy between gravitational masses and luminous masses in
galaxies.   

\begin{figure}
  \begin{center}
    \includegraphics[scale=.43]{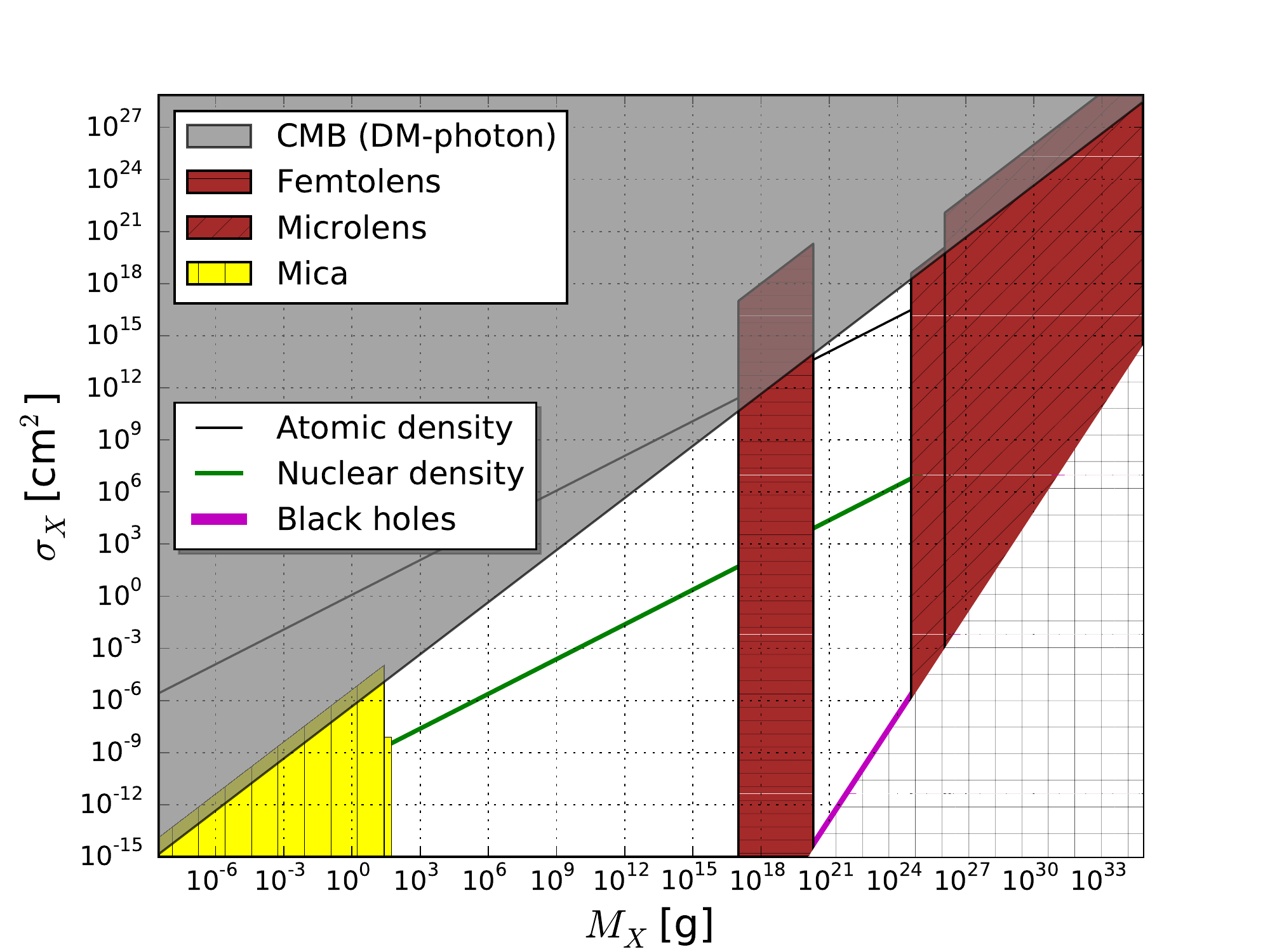}
  \end{center}
\caption{Figure 3 of \cite{Jacobs:2014yca}.  
\lq\lq{}Constraints on the macro cross section and mass 
(assuming the macros all have the same mass), 
applicable for both elastically- and inelastically-scattering candidates. 
In red are the femto- and micro-lensing constraints, 
while in grey are the CMB-inferred constraints (see \cite{Jacobs:2014yca} for references). 
The black and green lines correspond to objects of constant density 
$1\te{g}/\te{cm}^3$ and $3.6 \times 10^{14}\te{g}/\te{cm}^3$, respectively. 
Black hole candidates lie on the magenta line,
however, these may be ruled out for other reasons (see \cite{Jacobs:2014yca} for references);
objects within the hatched region in the bottom-right corner should not exist 
as they would simply be denser than black holes of the same mass.\rq\rq}

\label{model_independent}
\end{figure}

Observational limits on such macroscopic dark matter have been obtained
by several groups over the years.  
Some of these have been specific to the
original \lq\lq{strange matter}" paradigm, 
while others have been more phenomenological.  
Recently, one of us, with colleagues, presented a comprehensive
assessment of limits on such macros as a function of their mass and cross-section
\cite{Jacobs:2014yca}, identifying specific windows in that parameter space
that were as yet unprobed.  We later refined those in \cite{Jacobs:2015csa}.
An interesting window identified there was for macros with masses of greater
than about $55$g, and densities that included nuclear density.  
This is shown in Figure 3 of \cite{Jacobs:2014yca} 
reproduced here as Figure \ref{model_independent}.

In \cite{Jacobs:2014yca}, no mention was made of seismological bound on macros --
obtained by considering the effects of macros striking the Earth or Moon -- 
even though these could conceivably have intruded into the open window identified above $55$g.  
Here we justify that caution, 
reconsidering the seismological signals that could be observed
when such a dark matter candidate impinges on the  Moon, 
and finding that the signal
(and hence the limits on macro abundance) had been overestimated.

It has been suggested  \cite{banerdt2005seismic} 
that the energy deposition into the $1\,\te{Hz}$ range from a nuclearite (nuclear-density macro) 
impact on the Moon or Earth should be approximately $5\%$ of the total energy deposition. 
This, as shown in section II A, is a sizable overestimate 
even compared to our own purposefully generous estimate.
We produce a more accurate model of the seismic effects of macro impacts, 
including in our model the effects of geometric lensing, 
stratification of the Moon, 
anelastic attenuation, and geometric attenuation. 
The most important consideration is that even a sizable mass macro
is very small compared to the multi-kilometer wavelengths of seismic waves that propagate
unattenuated through the Moon or Earth. 
The production of detectable long-wavelength seismic waves is therefore highly suppressed
relative to short-wavelength waves.

For the balance of this paper, we focus on the Moon, rather than the Earth,
as the target system.
Unlike the Earth, the Moon is seismically quiet. 
Most internal seismological activity in the Moon originates from deep Moonquakes, 
which are very attenuated by the time they reach the surface. 
Most of the noise in lunar seismograms comes from meteoroids, 
but meteoroid impacts are for the most part easily identifiable as such. 
This allows us to place constraints on the macro flux from the total seismic event rate. 
A more sensitive search for macro impacts could seek, 
as Teplitz \textit{et al.} have \cite{banerdt2005seismic},
to identify the distinctive linear morphology of seismic events that a macro would cause 
as it bored straight through the target at high velocity (typically hundreds of km/s).
We look forward to that in a future paper.

The result of the current analysis is an upper bound on the event rate 
that would have been measured by the Apollo lunar seismometers
that substantially weakens the previously reported macro bound.
In particular, we find that macros of nuclear or greater density 
are not constrained by lunar seismic data.

\section{The Seismic Source}
\subsection{Seismic Wave Generation}

Once a  macro hits the Moon it can suffer a variety of fates.
If it is comparable in density to ordinary matter, 
then, like a  meteoroid, 
it will deposit all of its kinetic energy over a small distance 
and result in an impact crater at the lunar surface.
Here, we are more interested in macro candidates that are much denser, 
probably comparable to nuclear density, 
striking the Moon with an  impact speed characteristic 
of relative orbital speeds  in the Milky Way (several hundred km/s).
This is far in excess of the speed of sound  in rock, which is just a few km/s.

A precise evaluation of the seismic signal resulting from such an impact
would require detailed modeling of the response of lunar rocks to the passage of 
a hypersonic dense projectile.  
We will attempt below to put an upper limit on the strength of that seismic signal,
and thereby demonstrate that it is difficult to place strong limits on macros from lunar seismology.
We will therefore consistently {\em overestimate} the seismic signal.

Because of the hypersonic impact of the macro, 
we expect a good model of the initial effect of its passage through the rock to be
the instantaneous heating of all the rock in a column swept out
by the cross-section $\sigma_X$ of the macro. 
We approximate this as each atom of rock material acquiring a random velocity 
approximately equal in magnitude to the impact velocity of the macro.
The macro therefore deposits energy along a straight line through the Moon
at a rate
\begin{equation}
\td{E}{x}{} = \rho\sigma_X v_X^2\,.
\end{equation}
% \begin{equation}
% E_\te{initial} = \abs{\td{E}{x}{}}\ell = \rho\sigma_X v_X^2\ell\,.
% \end{equation}
A macro of mass $M_X$ will therefore traverse the Moon without significant deceleration if 
\begin{equation}
\label{eqn:nodeceleration}
2R_\te{Moon} \sigma_X \rho_\te{Moon}  \ll M_X\,,
\end{equation} 
where $\rho_\te{Moon}$ is the appropriately averaged lunar density 
and $2R_\te{Moon}$ is the lunar diameter.
For the balance of this paper, we will take equation (\ref{eqn:nodeceleration}) to apply,
unless specifically noted.

This energy deposition transforms the impacted rock into a column of ionized plasma,
and initiates an outward propagating melt front.  
We show in appendix \ref{app:Heat} below
that for the macro cross-sectional areas of interest to us,
the melt front either advances subsonically immediately, 
or at best the transition from supersonic to subsonic propagation
occurs marginally outside the macro radius.  
Once the velocity of the melt front becomes subsonic,
a seismic wave, sourced by the overpressure interior to the melt front, 
will begin to propagate radially outward ahead of the melt front.

Initially the outgoing wave will be non-linear, 
and will likely also result in the fracturing of the rock.
Eventually, the pressure amplitude of the wave will fall below $p_0$,
the maximum differential pressure  that the rock in this region can support elastically. 
From this time forward, we can treat the outgoing seismic wave 
as an ordinary linear wave.  

We begin by estimating the energy carried in that outgoing wave.
Taking the column of overpressured melted rock to be instantaneously generated, 
the resulting overpressure will be radially symmetric and of the form
\begin{equation}
p(r) = \left\{
\begin{array}{lr}
p_0 f(r)&\te{ if }a<r<b,
-\f \ell2<z<\f \ell2\,,\\ p = 0&\te{otherwise,}\end{array}
\right.
\end{equation}
where $0<f(r)<1$.
The overpressure extends over the entire length of the column ($-\ell/2<z<\ell/2$)
over some range of radii $a<r<b$.
After a short time, the pressure will remain in this cylindrically symmetric form
(at least away from the lunar surface), 
since the inhomogeneity of the Moon 
(and hence of the development of the evolving pressure distribution) 
is significant only on scales of tens or hundreds of kilometers.

The energy of a seismic $p$-wave is
\begin{equation}
\label{eqn:EnergySeismicWave}
E = \f12\int\diff^3 x\p{ \rho\vert{\partial_t u}\vert^2+(\lambda+2\mu)\vert{\nabla u}\vert^2}\,,
\end{equation}
where $u$ is the displacement field, 
and $\rho,\lambda,\mu$ are the density and Lam\`e coefficients respectively. 
For $p$-waves, $p = -K\nabla\cdot u$ where $K$ is the bulk modulus. 

We can rewrite (\ref{eqn:EnergySeismicWave}) in terms of the Fourier transform of $p(r)$.
To leading order in $k b$ this is
\begin{equation}
P = \f{4\pi p_0}{k\cos\theta}\sin\p{\f{h k\cos\theta}{2}}F+\scr O((kb)^0)\,,
\end{equation}
% where $\theta,\phi$ are the momentum-space polar angles, 
where $k$ is the magnitude of the wave-vector, 
$\theta$ is the angle the wave-vector makes with the tube axis,
and 
$F \equiv b^2f_1(b)-a^2f_1(a)$,
where
\begin{equation}
f_1(x) = 
\sum_{n=0}^\infty\sum_{m=0}^n\binom{n}{m}\f{f^{(n)}(x_0)}{n!}\f{(-x_0)^{n-m}}{2+m}x^{m}\,,
\end{equation}
for any choice of $x_0\in[a,b]$.
Defining $\kappa \equiv K^2/(\lambda + 2\mu)$, it follows that
\begin{equation}
\label{eqn:Etotal}
E =\f1\kappa\int\f{\diff^3 k}{(2\pi)^3}\vert{P}\vert^2\,.
\end{equation}

The seismometers that were left on the Moon by the Apollo astronauts, 
and which functioned until being decomissioned in 1977,
were sensitve at frequencies up to $20\,\te{Hz}$ 
to displacements as small as $0.3\,\text{nm}$ \cite{latham1973lunar,nakamura1982apollo}. 
The detectable energy is therefore
\begin{equation}
\begin{aligned}
E_k &= \f1\kappa\int_0^k\f{\diff k'\diff\theta}{(2\pi)^2}k'^2\sin\theta\vert{P}\vert^2\,\\
% \end{equation}
% Substituting in our $P$, we find that
% \begin{equation}
% \notag E_k
\notag&=\f{2p_0^2F^2k^2\ell }\kappa %\\\notag&\times
\left[\f{\sin(k\ell )+k\ell(\cos(k\ell)-2)}{k^2\ell^2 }+\te{Si}(k\ell )\right]\\
\notag&+\scr O((kb)^4)\,.
\end{aligned}
\end{equation}
When $k\ell >1$ 
\begin{equation}
E_k\simeq\f1\kappa \pi\ell p_0^2F^2 k^2\equiv C_1 k^2\,.
\end{equation}

The initial pressure profile is probably determined by the details of the initial 
macro-induced plasma, and the subsequent outward propagation of a melt front.
As described above,
once that front propagates out subsonically, 
a seismic p-wave will travel outward from the column of molten rock.
Initially, however, the overpressures in the wave
may be outside the linear regime for the rock elasticity.
As the pressure wave propagates outward, 
it is attenuated until the pressure differential is within the linear regime. 

The macro travels hypersonically and will generate a shock wave. 
Shock waves typically evolve to 
pressure fronts resembling a right triangle \cite{forbes2013shock}, 
i.e. $f = \p{r - r_0}/\Delta r$ where $a = r_0$ and $b = r_0 + \Delta r$. 
We adopt this shape also for its simplicity, in the expectation that
the specific shape is unlikely to grossly alter the conclusions. 
In this case $f_1(x) = \p{2 x-3r_0}/6\Delta r$ and
\begin{equation}
E_k = \f{\Delta r^2(3r_0+2\Delta r)^2}{36 \kappa} \pi\ell p_0^2k^2\,.
\end{equation}

The total energy of the pressure wave is (from (\ref{eqn:EnergySeismicWave}))
\begin{equation}
\begin{aligned}
E \notag&= \f{2\pi p_0^2\ell}{\kappa}\int_a^b r\diff r f(r)^2\,,\\
&=\f{p_0^2}{6\kappa}\pi\Delta r \ell(4r_0 + 3\Delta r)\,.
\end{aligned}
\end{equation}
Some of that energy would be lost to structural changes to the rock, 
such as melting and breaking.
The non-linear regime is characterized by faulting and fracturing. 
Brittle failure for granite occurs 
at a stress exceeding $3\times 10^8\,\te{Pa}$ \cite{lockner200232},
which is much less than the initial overpressure that the macros will 
leave behind in their melt tubes.
By ignoring dissipation during  the non-linear evolution of the seismic waves, 
we will, as intended,  overestimate the energy that would reach the seismometers. 
Overestimating the signal, 
we set $E = \rho\sigma_X \ell v_X^2\equiv\epsilon\ell$,
take $p_0 = \min\p{10^8\,\te{Pa},p_\te{source}}$,
% E_\te{propagated}$ 
and obtain an expression for $r_0$, which demarcates the end of the non-linear regime:
\begin{equation}
r_0 =  \f{3\kappa\epsilon}{2\pi p_0^2\Delta r}-\f{3\Delta r}{4}\,.
\end{equation} 
 
The fraction of the original deposited energy detectable to seismometers is thus
\begin{equation}
\Xi\equiv \f{E_k}{E} 
= k^2\f{\p{\pi p_0^2\Delta r^2-18\kappa \epsilon}^2}{576 \pi p_0^2 \kappa \epsilon}\,.
\end{equation}

A lower bound on $r_0$ is 0, 
corresponding to linear behavior from the start. 
(Actually, the lower bound is $\sqrt{\sigma_X/\pi})$, but the difference is negligible.)
This, in turn, imposes an upper bound on the pulse width
\begin{equation}
\Delta r\leq\sqrt{\f{2\kappa \epsilon}{\pi p_0^2}}\,.
\end{equation}
When $\Delta r$ is restricted to its physical range, 
$\Xi$ is a monotonic decreasing function in $\Delta r$. Thus
\begin{equation}
\f{4}{9}\f{\kappa\epsilon k^2}{\pi  p_0^2} \leq \Xi <\f{9}{16}\f{\kappa\epsilon k^2}{\pi  p_0^2}\,.
\end{equation}
It is important to note that these expressions only hold for $k(r_0+\Delta r)\ll 1$, 
however they will always provide an over-estimate of the fraction of detectable energy. 
Moreover, this condition holds for a wide variety of relevant parameters. 
Taking 
% $\epsilon_\te{melt} = 0$, 
$\rho = 3.3\times 10^3\,\te{kg m}^{-3}$, 
$v_X = 2.5\times 10^5\,\te{m s}^{-1}$, 
$\kappa = 5.5\times 10^{10}\,\te{Pa}$, 
and $k = 1.5\times 10^{-2}\,\te{m}^{-1}$ \cite{garcia2011very}, 
we obtain $\Xi<4.5\times (\sigma_X/\te{cm}^2)$. 
To obtain the $0.05$ that Teplitz \textit{et al.} used, 
one must take $\sigma_X >10^{-2}\te{\,cm}^2$, which is rather large 
\footnote{
	Note that $p_0\to 0$ does not imply $\Xi\to\infty$, 
	since $p_0$ is implicitly a function of $\epsilon$ when $p_0<10^8\,\te{Pa}$. 
	In this case, the ratio $\epsilon/p_0^2$ is always finite since $\epsilon\propto p_0^2$.
	}
(cf. Figure \ref{model_independent}).

The choice of $p_0 = 10^8\,\te{Pa}$ does significantly impact our result; 
ultimately the measured displacements we calculate will be inversely proportional to $p_0$. 
Our choice of $p_0$, however, 
is well below the overpressure corresponding to the boundary 
between linear and non-linear elasticity in the Moon. 
As the ambient pressure on a sample of rock increases, 
so too does $p_0$ \cite{mair2002influence,shimada1993lithosphere,ord1991deformation}.
We have taken $p_0$ to be a factor of $3$ 
below that of granite with a modest $50\te{ MPa}$ overpressure \cite{lockner200232}. 
In reality, $p_0$ is probably, on average, orders of magnitude larger than we claim it to be 
because the ambient pressure in the Moon is on the order of $\te{GPa}$ 
rather than $\te{MPa}$. 
This is one more way in which we overestimate the seismic signal.

To our knowledge, no measurements probe the elastic behavior of rock at these high pressures. 
Note also that this choice of $p_0$ corresponds to $r_0$ on the order of $1\te{ km}$, 
which well exceeds the regime of non-linearity we would have expected 
based on the solutions to the heat equation (Appendix \ref{app:Heat}).

\section{Seismic Wave Propagation}
\label{sec:SeismicPropagation}
The velocity of $p$-waves as a function of distance $r$ 
from the center of the  Moon 
(or the Earth, and presumably other spherical rocky celestial bodies) 
is of the form $v(r) = a^2 - b^2 r^2$ \cite{garcia2011very,dziewonski1981preliminary}
in each of a number of layers.
Using Snell's law, 
we obtain the differential equations for the trajectory of $p$-wave rays within each layer
\begin{equation}
\begin{aligned}
v^2 \notag&= \dot r^2 + \f{p_\te{ray}^2 v^4}{r^2}\,,\\
\dot\theta &=\pm\f{p_\te{ray} v^2}{r^2}\,.
\end{aligned}
\end{equation}
where $\theta$ is the polar angle of the ray measured from the center of the Moon, 
and $p_\te{ray}$ is the ray parameter, which is fixed along any given ray trajectory. 
These equations can be integrated with the above $v(r)$ to obtain, 
for some constants $q$ and $\theta_0$,
\begin{equation}
\begin{aligned}
\notag&r(t) = \f{a\sqrt{(qe^{2abt}-b)^2 + 4a^2b^4 p_\te{ray}}}{b\sqrt{(qe^{2abt}+b)^2 + 4a^2b^4 p_\te{ray}}}\,,\\
&\tan(\theta(t) - \theta_0) = \f{q^2e^{4abt}-b^2}{4ab^3p_\te{ray}}+abp_\te{ray}\,.
\end{aligned}
\end{equation}
%\textit{A priori} these are geodesics on the Poincar\'e disc of radius $\f ab$, which is easily seen from the form of $v$.

The  Moon, like the Earth, is stratified. 
At each boundary, a ray will be reflected and transmitted. 
We assume for simplicity 
that the reflection and transmission coefficients are frequency independent.
While this may  lead to either an overestimate or an underestimate of the signal,
the net effect of these reflections and transmissions on the limit is small,
and so their frequency dependence is unlikely to spoil the overall point.

The last effect to account for is anelastic attenuation (i.e. absorbtion), 
characterized by the quality factor $Q$, 
which depends on both frequencey and $r$ 
For a given mode, the ratio of final to initial amplitude is
\begin{equation}
\exp\ps{-k \int_{t_0}^t\diff t'\f{v(r(t'))}{2Q(r(t'))}}\equiv \exp\ps{-k {\cal D}}\,.
\end{equation}
The VPREMOON Model \cite{garcia2011very} 
provides piecewise-constant data for $Q$, 
so it is reasonable to subdivide the Moon further into strata of different $Q$. 
For propagation within a given layer $i$, 
the factor ${\cal D}$ is given by $\Delta t_i v_i/Q_i$, 
where we take $v_i$ to be the average velocity within that stratum. 
For the case of the  Moon (and the Earth), 
this is a good approximation, 
since $v$ doesn't change significantly within a given layer of constant $Q$. 
Again, while this approximation may lead 
to either an overestimate or an underestimate of the signal,
the net effect of the attenuation  on the limit is small,
because we are interested only in long wavelengthe modes which have less attenuation.

Thus, the amplitude of a ray can be computed by knowing the two numbers
\begin{equation}
\begin{aligned}
{\cal T} \notag&\equiv \prod_i{\cal T}_i\,,\\
{\cal D} &\equiv\sum_i\f{\Delta t_i v_i}{Q_i}\,,
\end{aligned}
\end{equation}
where ${\cal T}_i$ are the reflection or transmission coefficients  at boundary $i$ 
on which the ray is incident.

\section{Seismic Wave Detection}
Consider a $p$-wave traveling towards positive $x$, 
and with amplitude that is non-zero only within some region $S$ 
in the plane normal to to its motion. 
Denote $A \equiv \int_S\diff y\diff z$, 
and let $\chi_S$ be $1$ on $S$ and $0$ elsewhere. 
The displacement field of the wave is
\begin{equation}
u(\vec x,t) = \chi_{\R\times S}\int\f{\diff k}{2\pi}U(k)e^{-\I k(x-v_p t)},
\end{equation}
so
\begin{equation}
\label{eqn:EstimateAbove}
\vert{u(\vec x,t)}\vert\leq\int\f{\diff k}{2\pi}\vert{U(k)}\vert\,.
\end{equation}
Its energy is
\begin{equation}
\label{eqn:WaveEnergy}
E = \rho v_p^2 A\int\f{\diff k}{2\pi}k^2\vert{U(k)}\vert^2\,.
\end{equation}
As before, we denote the energy in the low frequency spectrum
\begin{equation}
E_k = \rho v_p^2 A\int_0^k\f{\diff k'}{2\pi}k'^2\vert{U(k')}\vert^2\,.
\end{equation}
It follows that
\begin{equation}
\vert{U(k)}\vert = \sqrt{\f{2\pi}{\rho v_p^2 A}}\f1k\sqrt{\td{E_k}{k}{}}\,,
\end{equation}
and from the estimate above (\ref{eqn:EstimateAbove})
\begin{equation}
\vert{u(\vec x,t)}\vert\leq(2\pi\rho v_p^2 A)^{-1/2}\int_0^k\f{\diff k'}{k'}\sqrt{\td{E_{k'}}{k'}{}}\,.
\end{equation}
In our case $E_k = \Xi E= C_1 k^2$ before attenuation. 
After anelastic attenuation, $d E_k/d k = 2{\cal T}C_1 k e^{-k\,{\cal D}}$
\begin{equation}
\vert{u(\vec x,t)}\vert\leq\sqrt{\f{2}{\rho v_p^2 A}}\sqrt{\f{{\cal T}\,\Xi E}{{\cal D}}}\f1k\te{Erf}\ps{\sqrt{\f{{\cal D}\,k}{2}}}\,.
\end{equation}
No similar simple analytic expression emerges
when $E_k$ includes the energies of two or more different rays, 
each with different attenuation factors.
If attenuation were important, 
we could average ${\cal D}$ among the coincident rays 
and obtain an approximate upper bound on  $\vert{u(\vec x,t)}\vert$.
However, since we only wish to consider the lowest frequency modes, $f<20\te{ Hz}$, 
and ${\cal D} k\ll1$, 
so absorption does not  significantly alter $u$. 
Taking $k{\cal D} = 0$, 
the displacement caused by the incident $p$-wave is bounded above by
\begin{equation}
\label{eqn:noattenuationumax}
\vert{u(\vec x,t)}\vert\leq\sqrt{\f{4{\cal T}\,\Xi E}{\pi k\rho v_p^2 A}}\,.
\end{equation}

\section{Simulation}
In previous works  (e.g. \cite{banerdt2005seismic}), 
homogeneous Earth and Moon models have been used 
to constrain the parameter space of macroscopic dark matter. 
This approach neglects the lensing that occurs because of the velocity gradient 
and the spherical boundaries of strata, 
as well as the energy losses from reflection and refraction across these boundaries. 
Here, we propagate rays, 
each carrying a fraction of the total energy deposited by the macro, 
from the macro's original line of impact to the boundary of the Moon. 
We then create an intensity map of the lunar surface 
for a representative sample of macro impacts. 
Finally, we convert the intensity maps to displacement maps, 
and using the sensitivity of the lunar seismometers, 
we obtain an average event rate that the lunar seismometers would measure.

\subsection{Data Generation}

We consider a macro trajectory that passes a distance $D$ from the lunar center. 
The trajectory has some length $L$ within the moon, which we sample at $M$ points. 
From each point we propagate $N$ randomly oriented rays, 
each endowed with energy $E_i = \rho(r_{i})\sigma_Xv_X^2 L/(MN)$, 
where $i\in\{1,\dots,M\}$ labels the points.
We use the trajectories derived in section \ref{sec:SeismicPropagation} to propagate the rays. 
Because of the approximation of spherical symmetry, each ray propagates in a plane. 

Since the moon is stratified, 
we split the propagation of a ray at the boundary of each layer 
into a reflected and a transmit ray.
During a ray's propagation through a given stratum, 
its time spent in that layer $\Delta t_i$, 
the attenuation factor in that layer ${\cal D}_i$, 
and the reflection/transmission coefficient ${\cal T}_i$ are all recorded. 
Each time a ray reaches the surface of the Moon, 
its position, cumulative propagation time $\Delta t$ 
and other trajectory attributes 
% (e.g.  ${\cal D}$ and ${\cal T}$) 
are recorded.
% The data produced by the propagation routine is a table of these values 
% and the corresponding surface coordinates 
% (i.e. we only count the energy of those rays which hit the surface), 
% as well as density at the source, $\kappa$ at the source, and $p$-wave velocity at the source. 
In this full model, it would take 16 iterations to propagate a ray 
from one side of the moon to the other. 
Since $Q$ is nearly constant, we reduce the number of boundaries by taking  $Q = 6750$. 
In this case, it takes 8 iterations to propagate through the Moon.
We expect this to lead to an overestimate of the signal, 
since each encounter with a boundary 
reduces the amplitude of the wave that reaches the surface.

\subsection{Data Analysis}

To analyze the simulations, 
we convert the surface-incident ray position data to HEALPix pixelization \cite{HEALPix}. 
(We use  $n_\te{side}=16$). 
The area of a HEALPix pixel is $A_\te{pix} = 4\pi R_\te{moon}^2/n_\te{pix}$,
where $n_\te{pix} = 12\times n_{\te{side}}^2$. 
A seismic wave can therefore cross a HEALPix pixel on the surface in 
$t_\te{crossing} \simeq \sqrt{A_\te{pix}/\pi v_\te{surface}^2}$. 
According to \cite{garcia2011very}, $v_\te{surface} \simeq 1\,\te{km s}^{-1}$. 
Rays that reach a surface pixel within $t_\te{crossing}$ of one another 
are taken to add constructively. 

We distribute the energy of each ray evenly over its pixel, 
and thus set $A$ (as in equation (\ref{eqn:WaveEnergy}) and following) 
equal to $A_\te{pix}$. 
This is a good approximation when $N=n_\te{pix}$, 
since the average angular separation of the rays at the source 
corresponds to the average angular separation of the HEALPix pixels. 
We omit the effects of anelastic attenuation,
as inspection of the propagation data showed 
that typically $k{\cal D}\simeq10^{-2}$ 
for the frequencies of interest in the Moon.

Using (\ref{eqn:noattenuationumax}), 
we compute an upper bound for the displacement within a given surface pixel, 
and compare that to the sensitivity of the lunar seismometers given in \cite{latham1973lunar}. 
If the calculated displacement is less than the seismometer sensitivity, 
then the seismic wave  is considered undetectable in that pixel. 
The number of detectable pixels 
is divided by the total number of pixels to obtain the fraction $\scr F_D$ 
of the lunar surface with a detectable signal 
for a macro making closest approach $D$ to the lunar center.

It is unlikely that a macro impact will be detected 
if it only just exceeds the seismometer sensitivity in one seismometer. 
As the displacement approaches $d_\te{min} = 3\times 10^{-10}\te{ m}$ in a pixel
with a seismometer, we should expect no chance of detection, 
while as $d\to\infty$ the impact should always be detected. 
We model this by taking the probability of detecting a seismic signal 
with displacement $\Delta x_i$
in the $i$th surface pixel (if it contains a seismometer) to be
\begin{equation}
m_{D,i} \equiv \max\pc{0,1-\exp\ps{p\p{1-\f{\Delta x_i}{d_{\te min}}}}}\,.
\end{equation}
Here $p$ describes how the detection probability increases as the strength of signal increases. 
(For example, if ${\Delta x}=4d_\te{min}$ 
is $q$ times as detectable as ${\Delta x}=2d_\te{min}$,
then $p = \ln\sqrt q$. )
Summing over all the pixels 
yields the effective number of pixels in which there exists a detectable signal
\begin{equation}
\label{eqn:singlepixeldetectionprobability}
m_D = \sum_{i = 1}^{n_{\te{pix}}}m_{D,i}\,.
\end{equation}
% \pc{1-\exp\ps{p\p{1-\f{\Delta x_i}{d}}}}\,.
In the end, we take $p\to\infty$ so as to, 
once again, overestimate the probability of detecting a lunar macro impact.

This procedure is repeated for $20$ values of the impact parameter $D$, 
evenly spaced  by $85\,\te{km}$. 
The average number of pixels on which there exists a detectable signal is then
\begin{equation}
m =\f{\sum_D D m_D}{\sum_D D}\,.
\end{equation}
If there are $n$ lunar seismometers (or tight clusters thereof), 
we suppose that each occupies one HEALPix pixel. 
If $n+m\geq n_\te{pix}$, then detection is guaranteed. 
Otherwise, if $\scr M$ macros impact the moon, 
the likelihood of any of them being detected is
\begin{equation}
\scr P = 1-\ps{\f{(n_\te{pix}-n)!(n_\te{pix}-m)!}{n_\te{pix}!(n_\te{pix}-n-m)!}}^{\scr M}\,.
\end{equation}
$\scr P$ is a function of $\sigma_X$, but not of $M_X$

% \subsection{Homogeneous Estimate}
% To obtain an estimate of the size of a detectable macro, 
% we suppose that one ray from each point source on the macro trajectory hits a single pixel. 
% In the case that rays don't bend, 
% this is a good approximation for behavior away from the source. 
% This approximation gives
% \begin{equation}
% u\leq\sqrt{\f{\Xi E}{\pi^2\rho k v_p^2 R_\te{Moon}^2}}\,.
% \end{equation}
% This expression is independent of the choice of $n_\te{side}$. 
% For the triangle wave, $\sigma_X\leq 1\times 10^{-7}\te{\,cm}^2$ 
% is unlikely to be detected unless it enters or exits the Moon adjacent to a seismometer.

\section{Results}

We have taken $p=\infty$ in (\ref{eqn:singlepixeldetectionprobability})
when calculating $m_D$, so as to overestimate the signal. 
The resulting detection probability versus macro cross-section 
is plotted in Figure 2.
%\ref{Detection_Probability}. %FOR SOME REASON THIS FIGURE REFERENCE ISN'T WORKING, AND I CAN'T FIGURE OUT WHY - it calls the figure VI, which makes no sense (although that's what section we're in)
 A very good fit to the curve is
\begin{equation}
\label{eqn:mDfit}
% \scr P(\sigma_X) \simeq\te{Erf}[(\alpha \sigma_X)^3 + (\beta\sigma_X)^2]
\scr P(\sigma_X) \simeq\te{Erf}[(\sigma_X/\sigma_0)^2(1+ \sigma_X/\sigma_1)]
\end{equation}
% with $\alpha=5.95\times 10^6\te{ cm}^{-2}$ and $\beta = 2.14\times 10^6\te{ cm}^{-2}$,
with $\sigma_0=4.67\times 10^{-7}\te{cm}^{-2}$ and $\sigma_1=2.17\times10^{-8}\te{cm}^{-2}$.
There is thus a very clear transition around $\sigma_X\simeq10^{-7}\te{ cm}^2$ 
where the detection probability increases rapidly with increasing cross-section.
% Not entirely surprisingly, this transition is close to the homogeneous
% estimate obtained above.

% Because macros are line sources, the exact seismic signature that would be seen at detectors is uncertain, so we use the total seismic event rate (NEED TO FILL OUT THE DISCUSSION HERE). This yields an exclusion curve (figure 3). 

Using the computed detection probability ${\scr P}	(\sigma_X)$, 
and the flux of macros impinging on the Moon 
\begin{equation}
\Phi(M_X)\simeq 5\times 10^7 (M_X/\te{ g})^{-1} \te{ yr}^{-1}\,,
\end{equation}
we can determine which values in the $\sigma_X$ vs. $M_X$ parameter plane
could be excluded using the Apollo seismometer data.
The reported rate of seismic events detected on the Moon by the four (clusters) 
of Apollo seismometers was $2500$ events per year.  
Therefore a good approximation of the excluded region is
\begin{equation}
\label{eqn:exclusion}
\scr P(\sigma_X) \Phi(M_X) > 2500.
\end{equation}
(If one had confidence that at most some fraction $f$ of these could be macros,
then for fixed $M_X$ one could improve the constraint on $\sigma_X$
but only by at most  $f^{1/2}$, due to the functional form of $\scr P(\sigma_X)$.)

In figure 4, we reproduce 
Figure 3 of \cite{Jacobs:2014yca} (i.e. Figure \ref{model_independent}) 
with the \lq\lq{exclusion region}\rq\rq~~ defined by (\ref{eqn:exclusion}) shaded blue.
For $M_X>2\times 10^4\te{g}$, 
the flux of macros is too low to give the observed seismic event rate;
below the region, the energy deposited is too small for the events to be detected.
Since our calculation assumed that the macro passed through the entire moon without
significant slowing,
it applies only for $M_X  > \sigma_X \rho_\te{Moon} 2R_\te{Moon}$,
which forms the upper boundary of the solid blue limit.
However, for fixed $M_X$, increasing $\sigma_X$ above $M_X/(\rho_\te{Moon} 2R_\te{Moon})$
seems unlikely to significantly reduce the signal, since the total energy deposited
will remain fixed but along a shorter path, nearer the lunar surface. We thus
expect the region above the solid blue region to be excluded as well, 
and cross hatch it in blue.

{\em We emphasize that since we have consistently overestimated the detectability of the signal,
the actual excluded region of parameter space lies above the shaded region
(i.e. at higher $\sigma_X$ for any given $M_X$). }

\begin{figure}
\label{Detection_Probability}
  \begin{center}
    \includegraphics[scale=0.42]{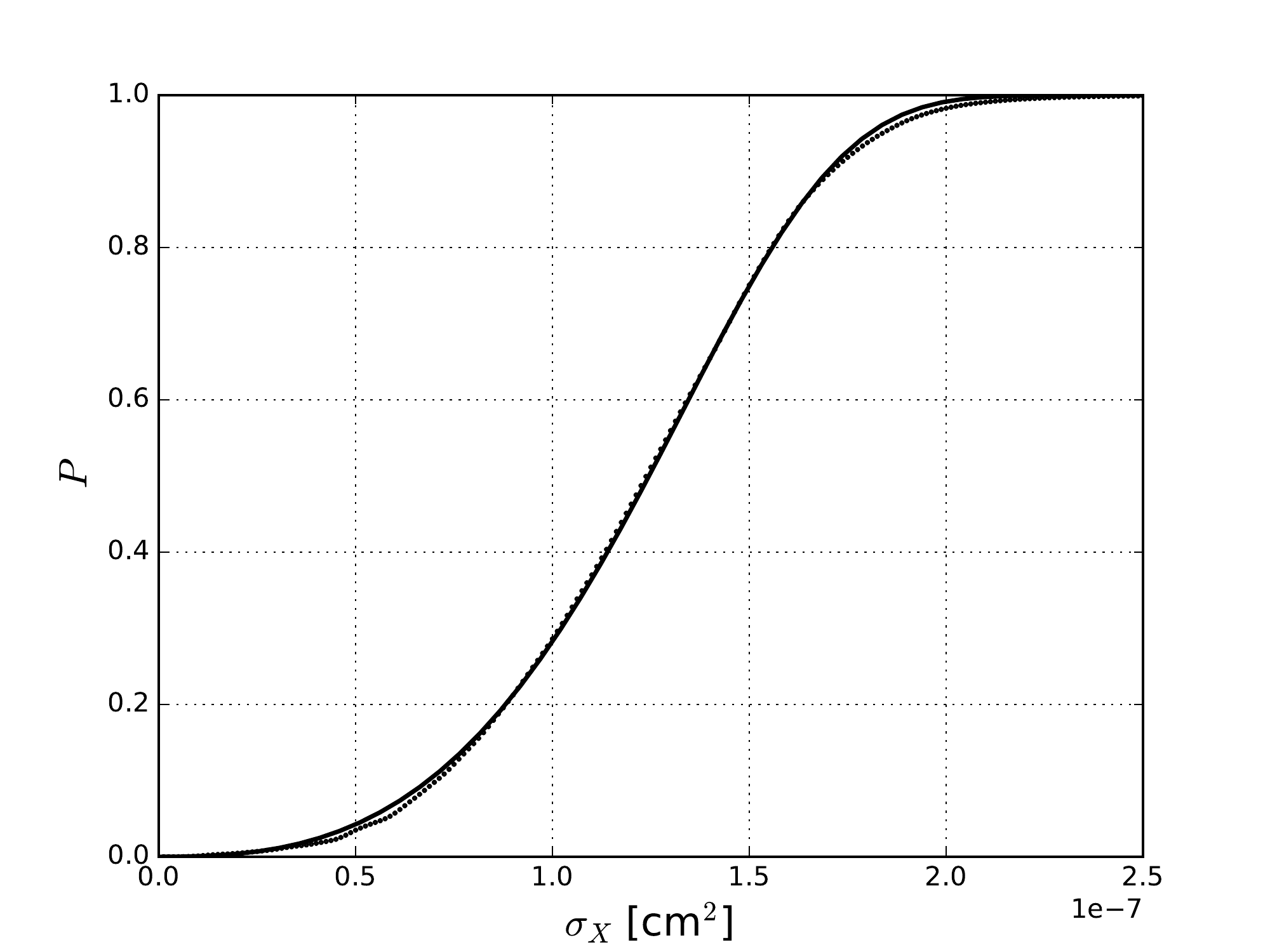}
  \end{center}
\caption{Detection Probability versus Cross-Section. 
The curve of best fit (solid) plotted is of the form 
$\te{Erf}[(\sigma_X/\sigma_0)^2(1+ \sigma_X/\sigma_1)]$
with $\sigma_0=4.67\times 10^{-7}\te{cm}^{-2}$ and $\sigma_1=2.17\times10^{-8}\te{cm}^{-2}$.}
\end{figure}
\begin{figure}
\label{Reproduce}
  \begin{center}
    \includegraphics[scale=0.43]{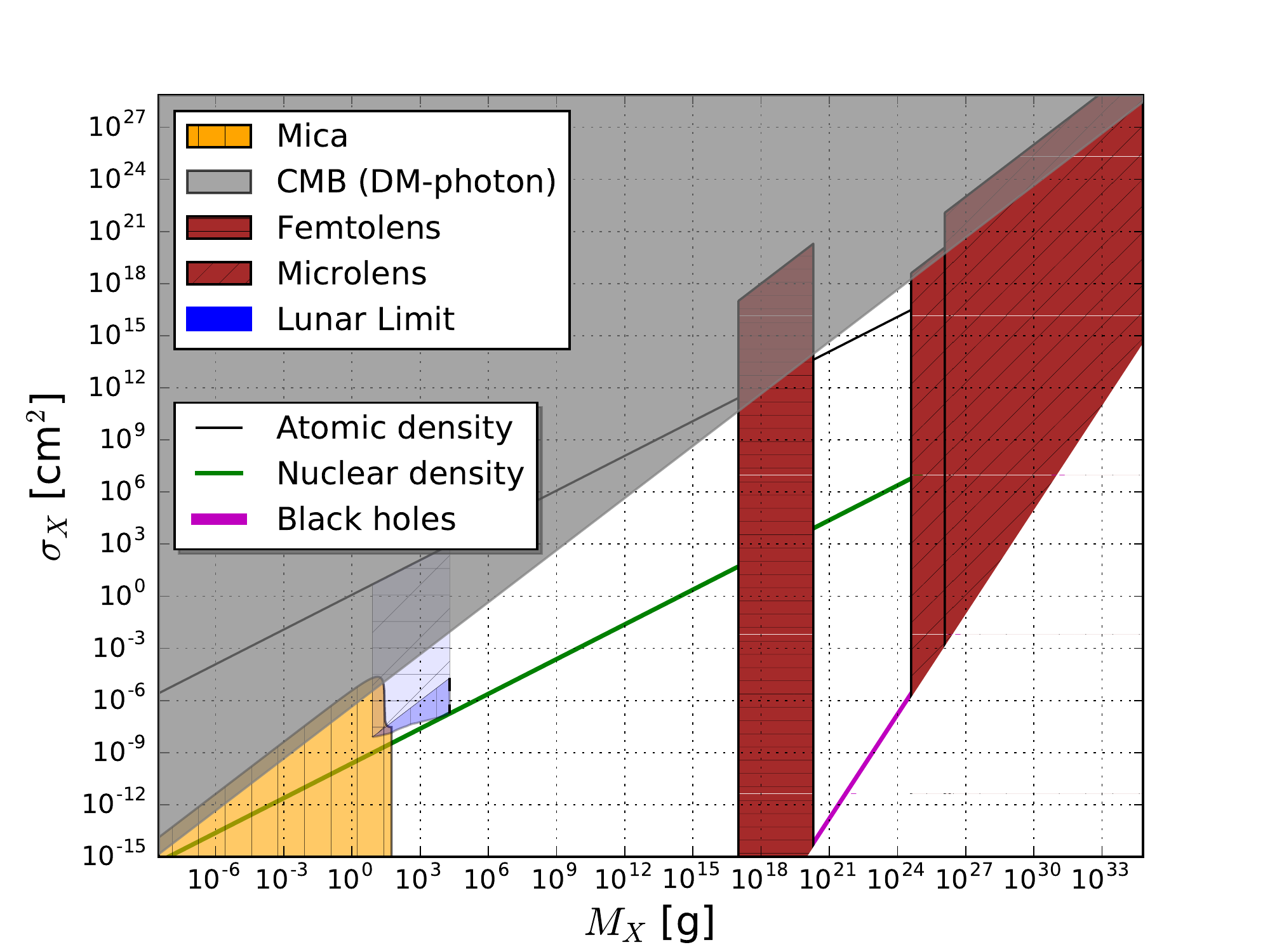}
  \end{center}
\caption{Figure 3 of \cite{Jacobs:2014yca} with the maximum seismic \lq\lq{excluded}\rq\rq~~ region added in blue. 
To the right of the blue region, the event rate is too low because the macro flux is too low; below the region, the signal strength is definitely too low. 
The excluded region is divided in two.  
In the solid blue colored region, our calculations apply.  
In the hatched blue region, 
the macro stops inside the moon and our calculations are not valid, 
but we expect this region is excluded as well. 
 We note that the nuclear-density line lies outside the excluded region. 
Since we have consistently overestimated the signal, 
the true exclusion area from total lunar seismic rate
must lie {\em above} the lower boundary of the blue shaded area. }
\end{figure}

\section{Conclusion}
We have considered more carefully than in the past the seismic signal generated
by a macroscopic dark matter candidate incident on the Moon.
We have found that the signal is weakened compared to prior expectations, 
in particular by the mismatch between the small size of the macro 
and the long wavelength of the modes that propagate largely unattenuated
(and  to which the lunar seismometers were sensitive).
For macroscopic dark matter of density greater than or equal to approximately nuclear density 
we have found that one cannot infer limits from the 
existing data on the total lunar seismological event rate.

The Earth and the Moon are ideal targets for looking for macros, 
whose vast surface area is difficult to improve upon.
However, in order to forecast the sensitivity of future seismological searches,
or to interpret future data, it will be necessary to markedly improve
our understanding of how, and with what efficiency, 
macro kinetic energy gets transferred into detectable seismic waves.

\acknowledgements
GDS thanks David Jacobs for the initial conversations regarding these seismological limits,
and the need to recompute the energy transfer from the macros to detectable seismic waves. 
GDS also thanks B. Lynn for early conversations about macros, 
and S. Iram for her contributions to preliminary investigations.
This work was partially supported by Department of Energy grant DE-SC0009946 
to the particle astrophysics theory group at CWRU. 
DC thanks SOURCE for partially supporting this research.
Some of the results in this paper have been derived using the HEALPix (\cite{HEALPix}) package. 
This work made use of the High Performance Computing Resource in the Core
Facility for Advanced Research Computing at Case Western Reserve University.
\appendix
\section{Propagation of the melt front}

\label{app:Heat}
When a macro impacts the Moon, 
the rock nearby is rapidly ionized, vaporized, or melted 
depending on its proximity to the macro. 
The energy deposition into these changes of phase 
will not contribute to primary seismic wave generation. 
The expansion of the radius from the macro trajectory to the boundary of the melted rock, 
the ``melt-front,'' slows down with time. 
When the melt-front velocity is below the speed of sound, 
seismic waves will escape the melted region, the ``melt-zone''. 
These seismic waves carry away  energy 
that has not already been used in phase transitions. 
	Because we are interested in an upper bound on the seismic activity of macro impacts, we assume that the all the remaining energy propagates away as seismic waves.

Nuclear dense macros have cross-section $\sigma_X$ small enough that they are reasonably approximated by a delta source. The $t=0$ temperature field is fixed by equating the heat energy with the macro energy
\begin{equation}
T(r,0) = \vert{\f{dE}{dx}}\vert\f{\sigma_X}{2\pi \rho \sigma_X c_p}\f{\delta(r)}{r} = \f{\sigma_{X} v_X^2}{2\pi c_p}\f{\delta(r)}{r}\,,
\end{equation}
where $\vert{\frac{dE}{dx}}\vert$, $v_X$, $c_p$, $\rho$ are the energy deposition, macro velocity, impacted material heat capacity and density respectively. The temperature field then evolves into a Gaussian in $r$,
\begin{equation}
T(r,t) = \f{\sigma_{X} v_X^2}{4\pi \alpha c_p}\f{e^{-\f{r^2}{4t\alpha}}}{t}\,.
\end{equation} 
Setting $T(r,t) = T_\te{melt}$ yields an expression for the melt-radius,
\begin{equation}
r_\te{m}(t) = \sqrt{4t\alpha\ln\ps{C/t}}\,,
\end{equation}
where
\begin{equation}
C = \f{v_X^2\sigma_X}{4 \pi c_p \alpha T_\te{melt}}\,.
\end{equation}
The melt-radius velocity is then
\begin{equation}
\dot r_\te{m}(t) =\sqrt{\f{\alpha}{t}}\ps{\p{\ln\ps{C/t}}^{1/2}- \p{\ln\ps{C/t}}^{-1/2}}\,.
\end{equation}
For sufficiently small $t\ll C$ it is a reasonable approximation to take
\begin{equation}
\dot r_\te{m}(t) \approx\sqrt{\f{\alpha}{t}\ln\ps{C/t}}\,.
\end{equation}
We can solve for $t$ such that $\dot r_\te{m} = v_p$, the speed of longitudinal waves, and hence determine the radius $r_\te{fm}$ at which the melt wave propagation slows to less than the speed of sound,
\begin{equation}
r_\te{fm} = A\, W\ps{Dr_X^2}\,,
\end{equation}
where $W$ is the Lambert $W$ function,
\begin{equation}
D = \frac{v_X^2}{c_pT_\te{melt}A^2}\hspace{0.2cm}\te{ and }\hspace{0.2cm}A = \frac{2\alpha}{v_p}\,.
\end{equation}

There is a critical value of $r_X$, 
$r_c\equiv- A\,W[-(A^2 D)^{-1}]$ (where one takes the lower branch of $W$)
above which this calculation yields $r_\te{fm}<\sqrt{\sigma_X/\pi}$, which is unphysical. 
For $r_X>r_c$, we should take $r_\te{fm}=\sqrt{\sigma_X/\pi}$.
For both granite ($\alpha\approx10^{-6}\te{m}^2/\te{s}$, 
	$T_\te{melt} = 1.49\times 10^3 \te{K}$ and $c_p = 1.05\times 10^3\te{J}/\te{kg K}$ 
	\cite{robertson1988thermal}) 
and limestone ($\alpha\approx10^{-6}\te{m}^2/\te{s}$, 
	$T_\te{melt} = 2.87\times10^3\te{K}$ and $c_p = 1.04\times 10^3\te{J}/\te{kg K}$ 
	\cite{oglesby2014deep}),
using $v_p = 8\times 10^3\te{m}/\te{s}$ from the VPREMOON model \cite{garcia2011very}, 
and  macro velocity $v_X = 2.5\times 10^5\te{m}/\te{s}$, we find
 $r_c \simeq 3 \times 10^{-9}\te{m}$.  
For nuclear-density macros of the allowed masses ($M_X>55\te{ g}$), 
$r_c\lesssim\sqrt{\sigma_X/\pi}$,
so the melt zone is negligible, 
though this might not be so for some higher density candidate macros.

Additional peculiarities with modeling the macro melt-zone are that the steep temperature gradient is not well approximated by the linear heat equation, 
and that the material properties of the rocks themselves are not constant in temperature. 
As we do not trust our model to accurately predict the energy lost, 
we overestimate the signal by assuming that the macro deposits all its energy into seismic waves. 
However, if we did use the results of this analysis, the energy loss it predicts is many orders of magnitude less than the total energy deposition, so it is inconsequential.
\bibliography{Article_Draft}{}

%merlin.mbs apsrev4-1.bst 2010-07-25 4.21a (PWD, AO, DPC) hacked
%Control: key (0)
%Control: author (72) initials jnrlst
%Control: editor formatted (1) identically to author
%Control: production of article title (-1) disabled
%Control: page (0) single
%Control: year (1) truncated
%Control: production of eprint (0) enabled
\begin{thebibliography}{20}%
\makeatletter
\providecommand \@ifxundefined [1]{%
 \@ifx{#1\undefined}
}%
\providecommand \@ifnum [1]{%
 \ifnum #1\expandafter \@firstoftwo
 \else \expandafter \@secondoftwo
 \fi
}%
\providecommand \@ifx [1]{%
 \ifx #1\expandafter \@firstoftwo
 \else \expandafter \@secondoftwo
 \fi
}%
\providecommand \natexlab [1]{#1}%
\providecommand \enquote  [1]{``#1''}%
\providecommand \bibnamefont  [1]{#1}%
\providecommand \bibfnamefont [1]{#1}%
\providecommand \citenamefont [1]{#1}%
\providecommand \href@noop [0]{\@secondoftwo}%
\providecommand \href [0]{\begingroup \@sanitize@url \@href}%
\providecommand \@href[1]{\@@startlink{#1}\@@href}%
\providecommand \@@href[1]{\endgroup#1\@@endlink}%
\providecommand \@sanitize@url [0]{\catcode `\\12\catcode `\$12\catcode
  `\&12\catcode `\#12\catcode `\^12\catcode `\_12\catcode `\%12\relax}%
\providecommand \@@startlink[1]{}%
\providecommand \@@endlink[0]{}%
\providecommand \url  [0]{\begingroup\@sanitize@url \@url }%
\providecommand \@url [1]{\endgroup\@href {#1}{\urlprefix }}%
\providecommand \urlprefix  [0]{URL }%
\providecommand \Eprint [0]{\href }%
\providecommand \doibase [0]{http://dx.doi.org/}%
\providecommand \selectlanguage [0]{\@gobble}%
\providecommand \bibinfo  [0]{\@secondoftwo}%
\providecommand \bibfield  [0]{\@secondoftwo}%
\providecommand \translation [1]{[#1]}%
\providecommand \BibitemOpen [0]{}%
\providecommand \bibitemStop [0]{}%
\providecommand \bibitemNoStop [0]{.\EOS\space}%
\providecommand \EOS [0]{\spacefactor3000\relax}%
\providecommand \BibitemShut  [1]{\csname bibitem#1\endcsname}%
\let\auto@bib@innerbib\@empty
%</preamble>
\bibitem [{\citenamefont {Rubin}\ \emph {et~al.}(1980)\citenamefont {Rubin},
  \citenamefont {Thonnard},\ and\ \citenamefont {Ford}}]{Rubin:1980zd}%
  \BibitemOpen
  \bibfield  {author} {\bibinfo {author} {\bibfnamefont {V.~C.}\ \bibnamefont
  {Rubin}}, \bibinfo {author} {\bibfnamefont {N.}~\bibnamefont {Thonnard}}, \
  and\ \bibinfo {author} {\bibfnamefont {W.~K.}\ \bibnamefont {Ford},
  \bibfnamefont {Jr.}},\ }\href {\doibase 10.1086/158003} {\bibfield  {journal}
  {\bibinfo  {journal} {Astrophys. J.}\ }\textbf {\bibinfo {volume} {238}},\
  \bibinfo {pages} {471} (\bibinfo {year} {1980})}\BibitemShut {NoStop}%
%%CITATION = ASJOA,238,471;%%
\bibitem [{\citenamefont {Lelli}\ \emph {et~al.}(2016)\citenamefont {Lelli},
  \citenamefont {McGaugh},\ and\ \citenamefont {Schombert}}]{Lelli:2016zqa}%
  \BibitemOpen
  \bibfield  {author} {\bibinfo {author} {\bibfnamefont {F.}~\bibnamefont
  {Lelli}}, \bibinfo {author} {\bibfnamefont {S.~S.}\ \bibnamefont {McGaugh}},
  \ and\ \bibinfo {author} {\bibfnamefont {J.~M.}\ \bibnamefont {Schombert}},\
  }\href@noop {} {\  (\bibinfo {year} {2016})},\ \Eprint
  {http://arxiv.org/abs/1606.09251} {arXiv:1606.09251 [astro-ph.GA]}
  \BibitemShut {NoStop}%
%%CITATION = ARXIV:1606.09251;%%
\bibitem [{\citenamefont {Witten}(1984)}]{Witten:1984rs}%
  \BibitemOpen
  \bibfield  {author} {\bibinfo {author} {\bibfnamefont {E.}~\bibnamefont
  {Witten}},\ }\href {\doibase 10.1103/PhysRevD.30.272} {\bibfield  {journal}
  {\bibinfo  {journal} {Phys. Rev.}\ }\textbf {\bibinfo {volume} {D30}},\
  \bibinfo {pages} {272} (\bibinfo {year} {1984})}\BibitemShut {NoStop}%
%%CITATION = PHRVA,D30,272;%%
\bibitem [{\citenamefont {Lynn}\ \emph {et~al.}(1990)\citenamefont {Lynn},
  \citenamefont {Nelson},\ and\ \citenamefont {Tetradis}}]{Lynn:1989xb}%
  \BibitemOpen
  \bibfield  {author} {\bibinfo {author} {\bibfnamefont {B.~W.}\ \bibnamefont
  {Lynn}}, \bibinfo {author} {\bibfnamefont {A.~E.}\ \bibnamefont {Nelson}}, \
  and\ \bibinfo {author} {\bibfnamefont {N.}~\bibnamefont {Tetradis}},\ }\href
  {\doibase 10.1016/0550-3213(90)90614-J} {\bibfield  {journal} {\bibinfo
  {journal} {Nucl. Phys.}\ }\textbf {\bibinfo {volume} {B345}},\ \bibinfo
  {pages} {186} (\bibinfo {year} {1990})}\BibitemShut {NoStop}%
%%CITATION = NUPHA,B345,186;%%
\bibitem [{\citenamefont {Jacobs}\ \emph
  {et~al.}(2015{\natexlab{a}})\citenamefont {Jacobs}, \citenamefont
  {Starkman},\ and\ \citenamefont {Lynn}}]{Jacobs:2014yca}%
  \BibitemOpen
  \bibfield  {author} {\bibinfo {author} {\bibfnamefont {D.~M.}\ \bibnamefont
  {Jacobs}}, \bibinfo {author} {\bibfnamefont {G.~D.}\ \bibnamefont
  {Starkman}}, \ and\ \bibinfo {author} {\bibfnamefont {B.~W.}\ \bibnamefont
  {Lynn}},\ }\href {\doibase 10.1093/mnras/stv774} {\bibfield  {journal}
  {\bibinfo  {journal} {Mon. Not. Roy. Astron. Soc.}\ }\textbf {\bibinfo
  {volume} {450}},\ \bibinfo {pages} {3418} (\bibinfo {year}
  {2015}{\natexlab{a}})},\ \Eprint {http://arxiv.org/abs/1410.2236}
  {arXiv:1410.2236 [astro-ph.CO]} \BibitemShut {NoStop}%
%%CITATION = ARXIV:1410.2236;%%
\bibitem [{\citenamefont {Jacobs}\ \emph
  {et~al.}(2015{\natexlab{b}})\citenamefont {Jacobs}, \citenamefont
  {Starkman},\ and\ \citenamefont {Weltman}}]{Jacobs:2015csa}%
  \BibitemOpen
  \bibfield  {author} {\bibinfo {author} {\bibfnamefont {D.~M.}\ \bibnamefont
  {Jacobs}}, \bibinfo {author} {\bibfnamefont {G.~D.}\ \bibnamefont
  {Starkman}}, \ and\ \bibinfo {author} {\bibfnamefont {A.}~\bibnamefont
  {Weltman}},\ }\href {\doibase 10.1103/PhysRevD.91.115023} {\bibfield
  {journal} {\bibinfo  {journal} {Phys. Rev.}\ }\textbf {\bibinfo {volume}
  {D91}},\ \bibinfo {pages} {115023} (\bibinfo {year} {2015}{\natexlab{b}})},\
  \Eprint {http://arxiv.org/abs/1504.02779} {arXiv:1504.02779 [astro-ph.CO]}
  \BibitemShut {NoStop}%
%%CITATION = ARXIV:1504.02779;%%
\bibitem [{\citenamefont {Banerdt}\ \emph {et~al.}(2005)\citenamefont
  {Banerdt}, \citenamefont {Chui}, \citenamefont {Herrin}, \citenamefont
  {Rosenbaum},\ and\ \citenamefont {Teplitz}}]{banerdt2005seismic}%
  \BibitemOpen
  \bibfield  {author} {\bibinfo {author} {\bibfnamefont {W.~B.}\ \bibnamefont
  {Banerdt}}, \bibinfo {author} {\bibfnamefont {T.}~\bibnamefont {Chui}},
  \bibinfo {author} {\bibfnamefont {E.~T.}\ \bibnamefont {Herrin}}, \bibinfo
  {author} {\bibfnamefont {D.}~\bibnamefont {Rosenbaum}}, \ and\ \bibinfo
  {author} {\bibfnamefont {V.~L.}\ \bibnamefont {Teplitz}},\ }in\ \href@noop {}
  {\emph {\bibinfo {booktitle} {The Identification of Dark Matter}}}\ (\bibinfo
  {year} {2005})\ pp.\ \bibinfo {pages} {581--586}\BibitemShut {NoStop}%
\bibitem [{\citenamefont {Latham}\ \emph {et~al.}(1973)\citenamefont {Latham},
  \citenamefont {Ewing}, \citenamefont {Dorman}, \citenamefont {Nakamura},
  \citenamefont {Press}, \citenamefont {Toks{\H{o}}z}, \citenamefont {Sutton},
  \citenamefont {Duennebier},\ and\ \citenamefont
  {Lammlein}}]{latham1973lunar}%
  \BibitemOpen
  \bibfield  {author} {\bibinfo {author} {\bibfnamefont {G.}~\bibnamefont
  {Latham}}, \bibinfo {author} {\bibfnamefont {M.}~\bibnamefont {Ewing}},
  \bibinfo {author} {\bibfnamefont {J.}~\bibnamefont {Dorman}}, \bibinfo
  {author} {\bibfnamefont {Y.}~\bibnamefont {Nakamura}}, \bibinfo {author}
  {\bibfnamefont {F.}~\bibnamefont {Press}}, \bibinfo {author} {\bibfnamefont
  {N.}~\bibnamefont {Toks{\H{o}}z}}, \bibinfo {author} {\bibfnamefont
  {G.}~\bibnamefont {Sutton}}, \bibinfo {author} {\bibfnamefont
  {F.}~\bibnamefont {Duennebier}}, \ and\ \bibinfo {author} {\bibfnamefont
  {D.}~\bibnamefont {Lammlein}},\ }\href@noop {} {\bibfield  {journal}
  {\bibinfo  {journal} {The Moon}\ }\textbf {\bibinfo {volume} {7}},\ \bibinfo
  {pages} {396} (\bibinfo {year} {1973})}\BibitemShut {NoStop}%
\bibitem [{\citenamefont {Nakamura}\ \emph {et~al.}(1982)\citenamefont
  {Nakamura}, \citenamefont {Latham},\ and\ \citenamefont
  {Dorman}}]{nakamura1982apollo}%
  \BibitemOpen
  \bibfield  {author} {\bibinfo {author} {\bibfnamefont {Y.}~\bibnamefont
  {Nakamura}}, \bibinfo {author} {\bibfnamefont {G.~V.}\ \bibnamefont
  {Latham}}, \ and\ \bibinfo {author} {\bibfnamefont {H.~J.}\ \bibnamefont
  {Dorman}},\ }in\ \href@noop {} {\emph {\bibinfo {booktitle} {Lunar and
  Planetary Science Conference Proceedings}}},\ Vol.~\bibinfo {volume} {13}\
  (\bibinfo {year} {1982})\ p.\ \bibinfo {pages} {117}\BibitemShut {NoStop}%
\bibitem [{\citenamefont {Forbes}(2013)}]{forbes2013shock}%
  \BibitemOpen
  \bibfield  {author} {\bibinfo {author} {\bibfnamefont {J.~W.}\ \bibnamefont
  {Forbes}},\ }\href@noop {} {\emph {\bibinfo {title} {Shock wave compression
  of condensed matter: a primer}}}\ (\bibinfo  {publisher} {Springer Science \&
  Business Media},\ \bibinfo {year} {2013})\BibitemShut {NoStop}%
\bibitem [{\citenamefont {Lockner}\ and\ \citenamefont
  {Beeler}(2002)}]{lockner200232}%
  \BibitemOpen
  \bibfield  {author} {\bibinfo {author} {\bibfnamefont {D.~A.}\ \bibnamefont
  {Lockner}}\ and\ \bibinfo {author} {\bibfnamefont {N.~M.}\ \bibnamefont
  {Beeler}},\ }\href@noop {} {\bibfield  {journal} {\bibinfo  {journal}
  {International Geophysics}\ }\textbf {\bibinfo {volume} {81}},\ \bibinfo
  {pages} {505} (\bibinfo {year} {2002})}\BibitemShut {NoStop}%
\bibitem [{\citenamefont {Garcia}\ \emph {et~al.}(2011)\citenamefont {Garcia},
  \citenamefont {Gagnepain-Beyneix}, \citenamefont {Chevrot},\ and\
  \citenamefont {Lognonn{\'e}}}]{garcia2011very}%
  \BibitemOpen
  \bibfield  {author} {\bibinfo {author} {\bibfnamefont {R.~F.}\ \bibnamefont
  {Garcia}}, \bibinfo {author} {\bibfnamefont {J.}~\bibnamefont
  {Gagnepain-Beyneix}}, \bibinfo {author} {\bibfnamefont {S.}~\bibnamefont
  {Chevrot}}, \ and\ \bibinfo {author} {\bibfnamefont {P.}~\bibnamefont
  {Lognonn{\'e}}},\ }\href@noop {} {\bibfield  {journal} {\bibinfo  {journal}
  {Physics of the Earth and Planetary Interiors}\ }\textbf {\bibinfo {volume}
  {188}},\ \bibinfo {pages} {96} (\bibinfo {year} {2011})}\BibitemShut
  {NoStop}%
\bibitem [{Note1()}]{Note1}%
  \BibitemOpen
  \bibinfo {note} {Note that $p_0\to 0$ does not imply $\Xi \to \infty $, since
  $p_0$ is implicitly a function of $\epsilon $ when $p_0<10^8\protect \tmspace
  +\thinmuskip {.1667em}\protect \text {Pa}$. In this case, the ratio $\epsilon
  /p_0^2$ is always finite since $\epsilon \propto p_0^2$.}\BibitemShut {Stop}%
\bibitem [{\citenamefont {Mair}\ \emph {et~al.}(2002)\citenamefont {Mair},
  \citenamefont {Elphick},\ and\ \citenamefont {Main}}]{mair2002influence}%
  \BibitemOpen
  \bibfield  {author} {\bibinfo {author} {\bibfnamefont {K.}~\bibnamefont
  {Mair}}, \bibinfo {author} {\bibfnamefont {S.}~\bibnamefont {Elphick}}, \
  and\ \bibinfo {author} {\bibfnamefont {I.}~\bibnamefont {Main}},\ }\href@noop
  {} {\bibfield  {journal} {\bibinfo  {journal} {Geophysical Research Letters}\
  }\textbf {\bibinfo {volume} {29}} (\bibinfo {year} {2002})}\BibitemShut
  {NoStop}%
\bibitem [{\citenamefont {Shimada}(1993)}]{shimada1993lithosphere}%
  \BibitemOpen
  \bibfield  {author} {\bibinfo {author} {\bibfnamefont {M.}~\bibnamefont
  {Shimada}},\ }\href@noop {} {\bibfield  {journal} {\bibinfo  {journal}
  {Tectonophysics}\ }\textbf {\bibinfo {volume} {217}},\ \bibinfo {pages} {55}
  (\bibinfo {year} {1993})}\BibitemShut {NoStop}%
\bibitem [{\citenamefont {Ord}(1991)}]{ord1991deformation}%
  \BibitemOpen
  \bibfield  {author} {\bibinfo {author} {\bibfnamefont {A.}~\bibnamefont
  {Ord}},\ }\href@noop {} {\bibfield  {journal} {\bibinfo  {journal} {Pure and
  Applied Geophysics}\ }\textbf {\bibinfo {volume} {137}},\ \bibinfo {pages}
  {337} (\bibinfo {year} {1991})}\BibitemShut {NoStop}%
\bibitem [{\citenamefont {Dziewonski}\ and\ \citenamefont
  {Anderson}(1981)}]{dziewonski1981preliminary}%
  \BibitemOpen
  \bibfield  {author} {\bibinfo {author} {\bibfnamefont {A.~M.}\ \bibnamefont
  {Dziewonski}}\ and\ \bibinfo {author} {\bibfnamefont {D.~L.}\ \bibnamefont
  {Anderson}},\ }\href@noop {} {\bibfield  {journal} {\bibinfo  {journal}
  {Physics of the earth and planetary interiors}\ }\textbf {\bibinfo {volume}
  {25}},\ \bibinfo {pages} {297} (\bibinfo {year} {1981})}\BibitemShut
  {NoStop}%
\bibitem [{\citenamefont {{G{\'o}rski}}\ \emph {et~al.}(2005)\citenamefont
  {{G{\'o}rski}}, \citenamefont {{Hivon}}, \citenamefont {{Banday}},
  \citenamefont {{Wandelt}}, \citenamefont {{Hansen}}, \citenamefont
  {{Reinecke}},\ and\ \citenamefont {{Bartelmann}}}]{HEALPix}%
  \BibitemOpen
  \bibfield  {author} {\bibinfo {author} {\bibfnamefont {K.~M.}\ \bibnamefont
  {{G{\'o}rski}}}, \bibinfo {author} {\bibfnamefont {E.}~\bibnamefont
  {{Hivon}}}, \bibinfo {author} {\bibfnamefont {A.~J.}\ \bibnamefont
  {{Banday}}}, \bibinfo {author} {\bibfnamefont {B.~D.}\ \bibnamefont
  {{Wandelt}}}, \bibinfo {author} {\bibfnamefont {F.~K.}\ \bibnamefont
  {{Hansen}}}, \bibinfo {author} {\bibfnamefont {M.}~\bibnamefont
  {{Reinecke}}}, \ and\ \bibinfo {author} {\bibfnamefont {M.}~\bibnamefont
  {{Bartelmann}}},\ }\href {\doibase 10.1086/427976} {\bibfield  {journal}
  {\bibinfo  {journal} {\apj}\ }\textbf {\bibinfo {volume} {622}},\ \bibinfo
  {pages} {759} (\bibinfo {year} {2005})},\ \Eprint
  {http://arxiv.org/abs/astro-ph/0409513} {astro-ph/0409513} \BibitemShut
  {NoStop}%
\bibitem [{\citenamefont {Robertson}(1988)}]{robertson1988thermal}%
  \BibitemOpen
  \bibfield  {author} {\bibinfo {author} {\bibfnamefont {E.~C.}\ \bibnamefont
  {Robertson}},\ }\href@noop {} {\emph {\bibinfo {title} {Thermal properties of
  rocks}}},\ \bibinfo {type} {Tech. Rep.}\ (\bibinfo  {institution} {US
  Geological Survey,},\ \bibinfo {year} {1988})\BibitemShut {NoStop}%
\bibitem [{\citenamefont {Oglesby}\ \emph {et~al.}(2014)\citenamefont
  {Oglesby}, \citenamefont {Woskov}, \citenamefont {Einstein},\ and\
  \citenamefont {Livesay}}]{oglesby2014deep}%
  \BibitemOpen
  \bibfield  {author} {\bibinfo {author} {\bibfnamefont {K.}~\bibnamefont
  {Oglesby}}, \bibinfo {author} {\bibfnamefont {P.}~\bibnamefont {Woskov}},
  \bibinfo {author} {\bibfnamefont {H.}~\bibnamefont {Einstein}}, \ and\
  \bibinfo {author} {\bibfnamefont {B.}~\bibnamefont {Livesay}},\ }\href@noop
  {} {\emph {\bibinfo {title} {Deep Geothermal Drilling Using Millimeter Wave
  Technology. Final Technical Research Report}}},\ \bibinfo {type} {Tech.
  Rep.}\ (\bibinfo  {institution} {Impact Technologies LLC, Tulsa, OK (United
  States)},\ \bibinfo {year} {2014})\BibitemShut {NoStop}%
\end{thebibliography}%
\bibliographystyle{apsrev4-1}

\end{document}